\documentclass{ifacconf}

\usepackage{graphicx}      
\usepackage{natbib}        
\usepackage[dvips]{epsfig}
\usepackage{amsmath}
\usepackage{bm}
\usepackage{amssymb}
\usepackage{amsfonts}
\usepackage{color}
\usepackage{subfigure}
\usepackage{algorithm}
\usepackage{algorithmicx}
\usepackage{circuitikz}

\newtheorem{Problem}{Problem}
\newtheorem{Remark}{Remark}

\newcommand{\sw}{{\bf s}}
\newcommand{\swest}{\hat{\bf s}}

\newcommand{\swseq}{{\bf S}}

\newcommand{\umeas}{{\bf u}}
\newcommand{\ymeas}{{\bf y}}
\newcommand{\yo}{{\bf y}^{\rm o} }

\newcommand{\NNparx}{\Theta_x}

\newcommand{ \NNpary}{\Theta_y}
\newcommand{\nin}{n_u} 
\newcommand{\ny}{n_y} 
\newcommand{\nx}{n_x} 
 
\newcommand{\Did}{\mathcal{D}} 
\newcommand{\nsamp}{N}

\newcommand{\hatxI}{\hat{\bf{x}}_I}
\newcommand{\hatx}{\hat{\bf{x}}}
\newcommand{\hatxdot}{\dot{\hat{{\bf{x}}}}}

\newcommand{\haty}{\hat{\bf{y}}}
\newcommand{\NN}{\mathcal{M}}

\newcommand{\hA}{\hat{A}}
\newcommand{\hB}{\hat{B}}
\newcommand{\hC}{\hat{C}}
\newcommand{\hD}{\hat{D}}

\newcommand{\Lcost}{\mathcal{L}}
\newcommand{\LcostInit}{\mathcal{L}^{\rm init}}
\newcommand{\LcostMode}{\mathcal{L}^{\rm mode}}
\newcommand{\LcostTrans}{\mathcal{L}^{\rm trans}}

\newcommand{\nmodes}{K}

\newcommand{\ba}[1]{\begin{array}{#1}}
	\newcommand{\ea}{\end{array}}

\newcommand{\smallmat}[1]{\left[ \begin{smallmatrix}#1 \end{smallmatrix} \right]}

\usepackage{enumitem}
\renewcommand{\theenumi}{\arabic{enumi}}

\renewcommand{\theenumii}{\arabic{enumii}}

\renewcommand{\theenumiii}{\arabic{enumiii}}

\begin{document}
\begin{frontmatter}

\title{Direct identification of continuous-time linear switched state-space models \thanksref{footnoteinfo1}} 

\thanks[footnoteinfo1]{This work has been submitted to IFAC World Congress'23 for possible publication.}

\author[First]{Manas Mejari} 
\author[First]{Dario Piga}

\address[First]{IDSIA Dalle Molle Institute for Artificial Intelligence, USI-SUPSI, Via la Santa 1, CH-6962 Lugano-Viganello, Switzerland. (e-mail: \{manas.mejari, dario.piga\}@supsi.ch).}

\begin{abstract}                
This paper presents an algorithm for direct  \emph{continuous-time} (CT) identification of \emph{linear switched state-space} (LSS) models. The key idea for direct CT identification is based on an integral architecture consisting of an LSS model followed by an integral block. This architecture is used to approximate the   continuous-time state map of a switched system.  A properly constructed  objective criterion is proposed based on the integral architecture in order to estimate the unknown  parameters and signals of the LSS model. A coordinate descent algorithm is employed to optimize this objective, which alternates between computing the unknown model matrices, switching sequence and estimating the state variables. 
The effectiveness of the proposed algorithm is shown via a simulation case study.
\end{abstract}


\end{frontmatter}

\section{Introduction}
\subsection{Linear switched systems}
Switched linear models belong to a class of hybrid systems, which consists of multiple linear subsystems and a switching signal dictating the \emph{active} linear subsystem at each time instance. Such model class is widely used to describe the behavior of dynamical systems subject to abrupt changes, exhibiting both continuous and  discrete dynamics. These changes can occur, for instance, due to sensor/actuator failures, external disturbances or a change in the operating point of a non-linear system. 

Over the past few decades, switched linear  models have found several applications in a variety of fields  including, mobile communication~\citep{abdollahi2011tcst}, signal processing~\citep{doucet2001tsp}, computer vision and bio-tracking~\citep{oh2008cv}, energy disaggregation~\citep{mnpb18cdc}, modeling human motion dynamics~\citep{pavlovic2000nips},   among many others.

\subsection{On direct continuous-time identification}

Concerning the identification of linear switched systems,  majority  of the  approaches proposed in the literature  have been developed for the identification of \emph{discrete-time} (DT) models. Among these, we mention optimization based algorithms~\citep{bako2011sparse, ohlsson2013auto}, recursive clustering-based approaches~\citep{BBPAut15,MeBrPi2020lcss},  mixed-integer programming algorithms~\citep{mnpb20ijrnc}, Bayesian inference~\citep{piga2020bayes}, algebraic-geometric approach~\citep{vidal2008auto}, 
which identify DT switched linear models in \emph{input-output} (IO) form. Although IO models are able to describe the behavior of the underlying system,  often it is desirable to estimate state-space representations,
as they are more convenient for stability analysis and controller synthesis of multi-input multi-output plants. To this end,  realization theory and subspace based algorithms have been developed for DT switched linear state-space models, see \citep{bako2009sysID,petreczky2013auto,verdult2004cdc}.  

In comparison with the large number of contributions  dedicated to the estimation of DT switched models, very few works have addressed direct \emph{continuous-time} (CT) identification of switched models. However, as discussed in \citep{garnier2015direct, Garnier:2008:ICM:1796449, PiCT18} for \emph{linear time-invariant} (LTI) models, direct  identification of CT model from sampled data offers multiple advantages over the discrete-time case. Most  of the physical systems are naturally modelled in continuous-time, and thus, the estimated parameters of CT models usually have a physical interpretation. Direct CT identification methods can also deal with non-uniformly sampled data, while discrete-time models implicitly rely on a fixed sampling time. Moreover, CT identification methods are generally more robust to numerical issues that may arise when using discrete-time methods in the case of high-frequency sampled data.

Motivated by these advantages, our goal in this paper is to develop an algorithm for direct CT identification for switched linear models. The core idea is based on the concept of \emph{integral architecture}, recently introduced by  the authors in \cite{MaFoPi20, mmfp22auto} for identification of CT non-linear and LPV systems. In this work, we extend that methodology for \emph{linear switched state-space} (LSS) model class.    

\subsection{Paper contributions and related works}

We consider the problem of direct CT identification of LSS models which involves:  estimating the matrices of each LTI submodel, computing a discrete mode sequence which indicates the active submodel at a given time, and estimating the continuous state-sequence, from a given sampled IO data. The proposed solution is based on  an integral architecture consisting of LSS model followed by an integral block, which is used to approximate the continuous state dynamics of an LSS system. A block coordinate descent algorithm is employed to optimize  a properly constructed dual-objective criterion, which alternates between computing the unknown matrices, discrete mode sequence and estimating the states. 

To the best of our knowledge, direct CT identification of switched linear models has been addressed very recently only in \citep{goudjil2020ecc, kersting2019ijc, du2021tcs}.      
These approaches, however, rely on strong assumptions imposed on the system's signals. In particular, the CT identification method proposed in \cite{goudjil2020ecc} requires that the input signal exciting the system is sinusoidal. Then, by exploiting the linearity of subsystems, outputs of individual subsystems are estimated using a DT switched IO method. In the next stage, CT identification approaches developed for LTI models are employed to estimate model parameters based on estimated subsystem outputs. In \cite{kersting2019ijc}, parameter identifiers and concurrent learning is proposed based on the assumptions that discrete mode sequence as well as continuous states are measured. Integral concurrent learning is proposed in~\cite{du2021tcs} relaxing the assumption of known switching sequence. However,  the continuous state is assumed to be measured.    
To position our work w.r.t. these contributions, we do not impose  any of the aforementioned assumptions required in \citep{goudjil2020ecc, kersting2019ijc, du2021tcs}, which are quite restrictive in practice.  
In particular,  in our contribution the input signals used to excite the system are not restricted to sinusoidal inputs. On the contrary, any class of input signals which excite all modes of the system can be used. Furthermore, neither the discrete mode sequence nor the continuous state are assumed to be known. The proposed algorithm estimates both these signals along with the model matrices through a block coordinate-descent approach  tailored to the considered identification problem. 

The  paper is organized as follows. The identification problem  for LSS models is formalized in Section~\ref{sec:ProblemSetting}.  The description of the integral architecture and details of proposed   proposed identification  algorithm are provided in Section~\ref{sec:CTI}. A simulation example is reported in Section~\ref{sec:CaseStudy}.

\section{PROBLEM FORMULATION}\label{sec:ProblemSetting}

We consider a  data-generating system $\mathcal{S}$ governed by the following CT  linear switched state-space representation:
\begin{subequations}
	\begin{align}
		\dot{\bf{x}}(t) &= A_{\sw(t)}{\bf{x}}(t) + B_{\sw(t)}{\umeas}(t),  \label{eq:init-a}\\
		{\bf{x}}(0) & = {\bf{x}}_0, \label{eq:init-b} \\
		{\yo}(t) & = C_{\sw(t)}{\bf{x}}(t) + D_{\sw(t)} \umeas(t)\label{eq:init-c},
	\end{align}
	\label{eq:init}
\end{subequations}%
where  ${\bf{x}}(t) \in  \mathbb{R}^{\nx}$ and $\dot{\bf{x}}(t) \in  \mathbb{R}^{\nx}$ are the state vector and its time derivative, respectively; ${\bf{x}}_0  \in  \mathbb{R}^{\nx}$ is the initial  condition;  ${\bf{u}}(t) \in  \mathbb{R}^{\nin}$ is the system input;  $\sw(t) \in \{1,\ldots,\nmodes \}$ is the switching signal and ${\yo}(t) \in  \mathbb{R}^{\ny}$ is the (noise-free) system output   at time $t \in  \mathbb{R}$. Note that, the system is assumed to have $K$ operating \emph{modes}, each corresponds to an LTI state-space system with real-valued matrices $\{A_i, B_{i},C_{i}, D_{i}\}_{i=1}^{\nmodes}$ of appropriate dimensions. 

A training dataset $\Did$  of length $\nsamp$ is gathered from the linear switched system $\mathcal{S}$ defined in \eqref{eq:init}  at time instants $\{t_0=0,t_1,\dots,\; t_{\nsamp-1} \}$. The dataset consists of input and \emph{noisy} output samples: $\Did= \{\umeas(t_k), \ymeas(t_k)\}_{k=0}^{\nsamp-1}$ with sampling time $\Delta t$.   
The measured output  is corrupted by a zero-mean white Gaussian noise $\eta \sim \mathcal{N}(0, \sigma^2_{\eta}I)$, \emph{i.e.}, $\ymeas(t_k) = \yo(t_k) + \eta(t_k)$.  

\begin{Problem}
	
	Given a training dataset $\Did$, our goal is to identify a continuous-time linear switched state-space (LSS) model, such that the model output  matches closely with the measured system output $\ymeas(t)$. 
\end{Problem}
The following assumptions are imposed on the system and signals in \eqref{eq:init}:
\begin{enumerate}
	 \item input signal ${\bf u}(t)$  can be  reconstructed (or reasonably approximated) for all time instants $t \in [0 \; \; t_{\nsamp-1}] \subset \mathbb{R}$ from the measured samples  $\{\umeas(t_k)\}_{k=0}^{\nsamp-1}$.
	\item time evolution of the switching signal  $\sw(t)$ is stochastic, but it's value does not change  during the sampling interval  $\Delta t$.
	\item all $K$ operating modes are distinguishable and sufficiently excited. 
\end{enumerate}

In the paper, we   consider a constant   sampling time $\Delta t$ only to ease the notation. Nevertheless, the approach is valid also for varying sampling times.

In the next paragraph, we introduce an \emph{integral architecture} for the identification of continuous-time LSS model.

\section{CONTINUOUS-TIME IDENTIFICATION OF LSS MODELS}
\label{sec:CTI}

\subsection{Integral architecture}
In order to describe the continuous-time state dynamics in \eqref{eq:init-a}, we define an LSS block  $\NN_x(\hatx,\umeas, \swest)$,  which is fed by the system input $\umeas(t)$, (estimated) switching signal $\swest(t)$ and  (estimated) state $\hatx(t)$ at time $t$, and returns the estimated state time-derivative  $\hatxdot (t)$, \emph{i.e.},   
\begin{align}
	&\NN_x(\hatx,\umeas,\swest):  \ \
	\hatxdot(t) \!=\! \hA_{\swest(t)} \hatx(t) +  \hB_{\swest(t)} \umeas(t),
	\label{eq:NNx}
\end{align}
where  $\hA_{i} \in \mathbb{R}^{\nx \times \nx}$ and $\hB_{i} \in  \mathbb{R}^{\nx \times \nin}$ (for $i=1,\ldots,\nmodes$) are the model matrices to be identified.  
Similarly, the output equation in \eqref{eq:init-c}  is represented by another block $\NN_y(\hatx,\umeas,\swest)$, which is fed by the estimated state $\hatx(t)$, input $\umeas(t)$ and estimated switching signal $\swest(t)$  and it returns the model output $\haty(t)$ at time $t$, \emph{i.e.},
\begin{align}
	&\NN_y(\hatx,\umeas,\swest): 
	\haty(t) \!= \hC_{\swest(t)} \hatx(t) + \hD_{\swest(t)} \umeas(t),
	\label{eq:NNy}
\end{align}
where the matrices $\hC_{i} \in \mathbb{R}^{\ny \times \nx}$ and $\hD_{i} \in  \mathbb{R}^{\ny \times \nin}$ (for $i=1,\ldots,\nmodes$) have to be estimated from data.

For brevity, we introduce the following notation: $\NNparx = \{\hA_{i}, \hB_{i} \}_{i=1}^{\nmodes}  $, $\NNpary = \{\hC_{i}, \hD_{i} \}_{i=1}^{\nmodes}  $ and the mode sequence $\swseq = \left(\swest(0),\ldots,\swest(t_{\nsamp-1})\right)$ 

The resulting continuous-time LSS model  is then given by:
\begin{subequations} \label{eq:final}
	\begin{align}
		\hatxdot(t) &= \NN_x(\hatx(t),\umeas(t), \swest(t); \NNparx(\swest(t))), \label{eq:final-a} \\
		\hatx(0)& =\hat{\bf{x}}_0, \label{eq:final-b}\\
		{\hat {\bf{y}}}(t) & = \NN_y(  \hatx(t),\umeas(t), \swest(t); \NNpary(\swest(t))), \label{eq:final-c}
	\end{align}
\end{subequations}
where $\NNparx(\swest(t))$ and $ \NNpary(\swest(t))$ denote the matrices $(\hA_{\swest(t)},$ $ \hB_{\swest(t)})$ and   $(\hC_{\swest(t)}, \hD_{\swest(t)})$ respectively, corresponding to the \emph{active mode} $\swest(t)$ at time $t$. 
Note that, for the $i$-the active mode, \emph{i.e.,} $\swest(t)=i$ at time $t$, the maps  $\NN_x(\cdot)$ and $\NN_y(\cdot)$ are \emph{linear} functions of the model matrices   $(A_i, B_i)$ and $(C_i, D_i)$ respectively.

In this paper, we adopt a method originally introduced  in \cite{MaFoPi20}, which exploits the \emph{integral form} of the Cauchy problem \eqref{eq:final-a}-\eqref{eq:final-b}, by defining an \emph{integral}  block $\NN_I$ as:
\begin{align} \label{eq:Ni}
	\hatxI(t) =  \NN_I(\hatx(t),\umeas(t), \swest(t); \NNparx(\swest(t))) 
\end{align}
with 
\begin{align*}
	&\NN_I(\hatx(t),\umeas(t), \swest(t); \NNparx(\swest(t))) \\&=  \hatx(0) + \int_{0}^{t}  \NN_x(\hatx(\tau),\umeas(\tau),  \swest(\tau);  \NNparx(\swest(\tau)))  d\tau.
\end{align*}
The block diagram in Fig.~\ref{fig:scheme_Integral} is a representation of  \eqref{eq:Ni}, along with the output equation \eqref{eq:final-c} producing ${\hat {\bf{y}}}(t)$.
Given the mode sequence $\swseq$, if the state $\hatx(t)$ feeding the LSS model block $\NN_I(\cdot)$ is actually generated by the model given in \eqref{eq:final}, then the state $\hatxI(t)$ exactly matches $\hatx(t)$, i.e., 
\begin{equation}
	\hatx(t) = \hatxI(t) \qquad \forall t \in [t_0\; t_{\nsamp-1}].
	\label{eq:xI_equal_x}
\end{equation}
\tikzset{every picture/.style={line width=0.75pt}} 
\begin{figure}
	\centering
	\begin{tikzpicture}[x=0.75pt,y=0.75pt,yscale=-1.2,xscale=1.2]
		\draw   (77,38) -- (125,38) -- (125,73) -- (77,73) -- cycle ;
		\draw   (150,38) -- (201,38) -- (201,73) -- (150,73) -- cycle ;
		\draw   (219.43,54.79) .. controls (219.43,51.04) and (222.47,48) .. (226.21,48) .. controls (229.96,48) and (233,51.04) .. (233,54.79) .. controls (233,58.53) and (229.96,61.57) .. (226.21,61.57) .. controls (222.47,61.57) and (219.43,58.53) .. (219.43,54.79) -- cycle ;
		\draw    (24,46.57) -- (74,46.22) ;
		\draw [shift={(77,46.2)}, rotate = 539.63] [fill={rgb, 255:red, 0; green, 0; blue, 0 }  ][line width=0.08]  [draw opacity=0] (5.36,-2.57) -- (0,0) -- (5.36,2.57) -- (3.56,0) -- cycle    ;
		\draw   (21,112) -- (69,112) -- (69,142.57) -- (21,142.57) -- cycle ;
		\draw    (125,56.2) -- (147.5,56.11) ;
		\draw [shift={(150.5,56.1)}, rotate = 539.78] [fill={rgb, 255:red, 0; green, 0; blue, 0 }  ][line width=0.08]  [draw opacity=0] (5.36,-2.57) -- (0,0) -- (5.36,2.57) -- (3.56,0) -- cycle    ;
		\draw    (24,55.57) -- (74,55.22) ;
		\draw [shift={(77,55.2)}, rotate = 539.63] [fill={rgb, 255:red, 0; green, 0; blue, 0 }  ][line width=0.08]  [draw opacity=0] (5.36,-2.57) -- (0,0) -- (5.36,2.57) -- (3.56,0) -- cycle    ;
		\draw    (24,64.57) -- (74,64.22) ;
		\draw [shift={(77,64.2)}, rotate = 539.63] [fill={rgb, 255:red, 0; green, 0; blue, 0 }  ][line width=0.08]  [draw opacity=0] (5.36,-2.57) -- (0,0) -- (5.36,2.57) -- (3.56,0) -- cycle    ;
		\draw    (233,54.79) -- (265,55.16) ;
		\draw [shift={(268,55.2)}, rotate = 180.68] [fill={rgb, 255:red, 0; green, 0; blue, 0 }  ][line width=0.08]  [draw opacity=0] (5.36,-2.57) -- (0,0) -- (5.36,2.57) -- (3.56,0) -- cycle    ;
		\draw    (201,55.2) -- (216.43,54.85) ;
		\draw [shift={(219.43,54.79)}, rotate = 538.71] [fill={rgb, 255:red, 0; green, 0; blue, 0 }  ][line width=0.08]  [draw opacity=0] (5.36,-2.57) -- (0,0) -- (5.36,2.57) -- (3.56,0) -- cycle    ;
		\draw    (50,46.39) -- (49.52,109) ;
		\draw [shift={(49.5,112)}, rotate = 270.44] [fill={rgb, 255:red, 0; green, 0; blue, 0 }  ][line width=0.08]  [draw opacity=0] (5.36,-2.57) -- (0,0) -- (5.36,2.57) -- (3.56,0) -- cycle    ;
		\draw    (40,55) -- (40,109.2) ;
		\draw [shift={(40,112.2)}, rotate = 270] [fill={rgb, 255:red, 0; green, 0; blue, 0 }  ][line width=0.08]  [draw opacity=0] (5.36,-2.57) -- (0,0) -- (5.36,2.57) -- (3.56,0) -- cycle    ;
		\draw    (30,64.2) -- (30,109.2) ;
		\draw [shift={(30,112.2)}, rotate = 270] [fill={rgb, 255:red, 0; green, 0; blue, 0 }  ][line width=0.08]  [draw opacity=0] (5.36,-2.57) -- (0,0) -- (5.36,2.57) -- (3.56,0) -- cycle    ;
		\draw  [dash pattern={on 4.5pt off 4.5pt}] (62,20) -- (239,20) -- (239,95.2) -- (62,95.2) -- cycle ;
		\draw    (226,26) -- (226.19,45) ;
		\draw [shift={(226.21,48)}, rotate = 269.44] [fill={rgb, 255:red, 0; green, 0; blue, 0 }  ][line width=0.08]  [draw opacity=0] (5.36,-2.57) -- (0,0) -- (5.36,2.57) -- (3.56,0) -- cycle    ;
		\draw    (45,143) -- (45,162) ;
		\draw [shift={(45,165)}, rotate = 270] [fill={rgb, 255:red, 0; green, 0; blue, 0 }  ][line width=0.08]  [draw opacity=0] (5.36,-2.57) -- (0,0) -- (5.36,2.57) -- (3.56,0) -- cycle    ;
		
		\draw (155,39) node [anchor=north west][inner sep=0.75pt]  [font=\small]  {$\int \limits_{0}^{t} \hatxdot(\tau) d\tau $};
		\draw (78.5,48) node [anchor=north west][inner sep=0.75pt]  [font=\small]  {$\mathcal{M}_{x}\left(\NNparx\right)$};
		\draw (23,120) node [anchor=north west][inner sep=0.75pt]  [font=\small]  {$\mathcal{M}_{y}\left(\NNpary\right)$};
		\draw (212,24.4) node [anchor=north west][inner sep=0.75pt]  [font=\small]  {${\bf{x}}_0$};
		\draw (245,39) node [anchor=north west][inner sep=0.75pt]  [font=\small]  {$\hatxI(t)$};
		\draw (1,40) node [anchor=north west][inner sep=0.75pt]  [font=\small]  {$\hatx(t)$};
		\draw (1,50) node [anchor=north west][inner sep=0.75pt]  [font=\small]  {$\umeas(t)$};
		\draw (1,60) node [anchor=north west][inner sep=0.75pt]  [font=\small]  {$\swest(t)$};
		\draw (127,39.5) node [anchor=north west][inner sep=0.75pt]  [font=\small]  {$\hatxdot (t)$};
		\draw (191,75.4) node [anchor=north west][inner sep=0.75pt]  [font=\small]  {$\mathcal{M}_{I}( \cdot ,\cdot )$};
		\draw (48,150.4) node [anchor=north west][inner sep=0.75pt]  [font=\small]  {$\haty(t)$};
		\draw (214,37.4) node [anchor=north west][inner sep=0.75pt]  [font=\tiny]  {$+$};
		\draw (209,44.4) node [anchor=north west][inner sep=0.75pt]  [font=\tiny]  {$+$};
	\end{tikzpicture}
	\caption{Integral architecture for continuous-time LSS model identification.}\label{fig:scheme_Integral}
\end{figure}
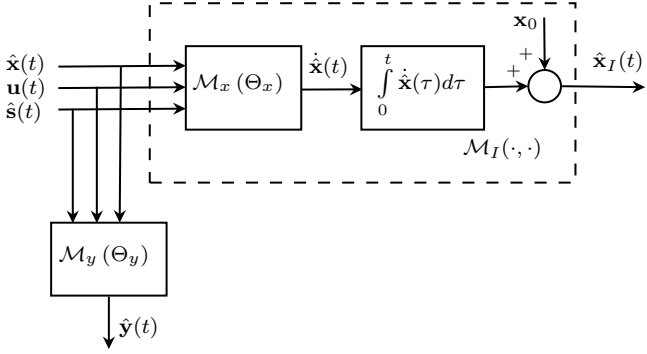

\subsection{Fitting criterion}

In the proposed scheme, the LSS model matrices  $\NNparx, \NNpary$, the mode sequence $\swseq$ and the state signal $\hatx(t),  t\in [t_0\;t_{\nsamp-1}]$, are \emph{free} optimization parameters. They are concurrently optimized according to a cost function  constructed  with the following rationale. 

First, the estimated model output $\hat{\bf{y}}$ should match the output measurements in training dataset $\Did$. This objective is achieved by introducing  a \emph{fitting term} $J_{\bf y}$  in the cost function 
which penalizes the mismatch between the model outputs $\haty(t_k)$ and the sampled measured outputs $\ymeas(t_k), \;k=0,1,\dots,\nsamp\!-\!1$.  

Second, the state signal $\hatx$ should be compatible with the LSS model dynamics \eqref{eq:final}. This can be achieved through an   additional \emph{regularization term} $J_{\bf x}$,  which  penalizes the distance between $\hatxI(t)$ and $\hatx(t)$, where $\hatxI$ is defined as in \eqref{eq:Ni}. The regularization term enforces the state $\hatx$ (which is to be optimized) to follow the CT model dynamics~\eqref{eq:final}.

Furthermore, a \emph{mode loss} term  $\Lcost: \mathcal{K}^{N} \rightarrow \mathbb{R} $ (where $\mathcal{K} = \{1,\ldots, \nmodes \}$) is imposed on the mode sequence $\swseq$  to take into account the temporal order as well as to incorporate the information of the switching mechanism, for \emph{e.g.}, Markovian switching in jump Markov linear systems  etc.

The following minimization problem is thus formulated:
\begin{subequations} \label{eqn:optprob}
	\begin{equation}
		\label{eq:optimization_problem}
		\begin{split}
			\underset{\hatx(\cdot), \swseq, \NNparx,\NNpary}{\text{min}} J(\hatx(\cdot),\swseq,\NNparx,\NNpary),
		\end{split}
	\end{equation}
	where
	\begin{equation}
		\begin{split} \label{eqn:J}
			J &= \underset{J_{\bf y}}{\underbrace{\sum_{k=0}^{N-1} \lVert  {\hat{\bf y}}(t_k)- { \bf y}(t_k)  \rVert^2 }} \\
			&+
			\alpha \underset{J_{\bf x}}{\underbrace{\int_{t_0}^{t_{N-1}} \lVert \hatxI(\tau)- \hatx(\tau) \rVert^2 \; d\tau}}
			+ \beta \ \Lcost(\swseq),
		\end{split}
	\end{equation}
	with 
	\begin{align} 
		\label{eq:output_equation_optim}
		\haty(t_k) & = \NN_y(\hatx(t_k),\umeas(t_k),\swest(t_k);\; \NNpary(\swest(t_k))), \\
		\label{eq:xI_optim}
		\hatxI(t) &=   \hatx(0) + \int_{0}^{t} \NN_x(\hatx(\tau),\umeas(\tau),\swest(\tau);\; \NNparx(\swest(t_k))) \;d\tau.
	\end{align}
\end{subequations}

As introduced in \citep{BPBB2018}, the following mode loss $\Lcost: \mathcal{K}^{N} \rightarrow \mathbb{R} $ can be considered 
\begin{align}\label{eq:mode_loss}
	\Lcost(\swseq) &= \LcostInit(\swest(0)) + \sum_{k=1}^{N-1}\LcostMode(\swest(t_k)) \nonumber \\
	&+ \sum_{k=1}^{N-1} \LcostTrans(\swest(t_k), \swest(t_{k-1})), 
\end{align}
where $\mathcal{K} = \{1,\ldots, \nmodes \}$,  $\LcostInit: \mathcal{K} \rightarrow \mathbb{R}$ is the initial mode cost,   $\LcostMode: \mathcal{K} \rightarrow \mathbb{R}$ is the mode cost, and $\LcostTrans: \mathcal{K}^{2} \rightarrow \mathbb{R}$ is the mode transition cost.

The hyper-parameters $\alpha, \beta > 0$ act as a tuning knob balancing the relative importance of the fitting cost $J_y$, the regularization cost  $J_x$ and the \emph{mode loss} $\mathcal{L}(\swseq)$. Additionally, in order to enforce smoothness properties for the estimated state variables, an $\ell_1$ regularization term $\|\hatx(t_k) -\hatx(t_{k-1}) \|_{1}$ can be also included in the optimization problem \eqref{eq:optimization_problem}.

\subsection{Integral approximation}

Note that the  continuous-time state signal $\hatx(t) \in \mathbb{R}^{\nx},\; t \in [t_0\;t_{\nsamp-1}]$ is one of the problem's decision variables.  Indeed, the optimization problem \eqref{eqn:optprob} is  \emph{infinite-dimensional} and thus computationally intractable. Following the rationale in \citep{mmfp22auto}, we employ numerical techniques  to approximate  \eqref{eqn:optprob} into a finite-dimensional problem amenable for a tractable implementation.
In particular, the state signal $\hatx(t)$ is approximated using a finite-dimensional parameterization. 
For simplicity of exposition, we represent the state signal with a 
\emph{piecewise constant} parameterization, where 
$\hatx(t)$ is constant during  the sampling intervals $[t_{k-1}\  t_{k}]$, $k=0,1,\dots,\nsamp-1$. In general, more complex parametrizations for $\hatx$ such as piecewise linear or polynomial could be also used. 
Moreover, the intervals for the piecewise constant approximation of $\hatx$ may not necessarily correspond to the input/output sampling time $\Delta t$  or the \emph{dwell time} of the switching signal.

Furthermore, we approximate the integrals in \eqref{eqn:J} and \eqref{eq:xI_optim} by applying a numerical integration scheme.
For  simplicity, in this work we apply the classical \emph{rectangular approximation} rule  for the numerical integration of \eqref{eqn:J} and \eqref{eq:xI_optim}. Other quadrature rules such as trapezoidal  or Gaussian quadrature could be alternatively considered.

Overall, the \emph{piecewise constant} parametrization of the signals 
$\hatx(t)$, $\umeas(t)$, $\swest(t)$  with the \emph{rectangular quadrature} of the integrals leads to the following approximation:

\begin{align*} 
	& \int_{t_0}^{t_{N-1}} \!\! \lVert \hatxI(\tau)- \hatx(\tau) \rVert^2 d\tau\!\approx \! \sum_{k=1}^{N-1} \lVert \hatxI(t_k)-\hatx(t_k) \rVert^2 \Delta t_k,  
\end{align*}
where $\Delta t_k=t_k-t_{k-1}$, and~\eqref{eq:xI_optim} can be approximated with the following Riemann sum:
\begin{align} \label{eqn:state-approx-2}
	\hatxI(t_k) &\approx \! \hat{\bf{x}}(0) \!+\! \sum_{j=0}^{k-1}   \Delta t_{j+1} \NN_x(\hatx(t_j),\umeas(t_j),\swest(t_j);\; \NNparx(\swest(t_j))). \\
	&= \hat{\bf{x}}(0) + \sum_{j=0}^{k-1}   \Delta t_{j+1}\left(\hA_{\swest(t_j)}\hatx(t_j)+\hB_{\swest(t_j)}\umeas(t_j)\right)
\end{align}

The sum in the equation above  can be also constructed recursively as follows:
\begin{align} \label{eqn:state-approx-3}
	\hatxI(t_k\!+\!1)\! =\! \hatxI(t_{k}) + 
	\overbrace{\Delta t_{k+1} \left(\hA_{\swest(t_k)}\hatx(t_k)+\hB_{\swest(t_k)}\umeas(t_k)\right) }^{\Delta {\bf{x}}_k}.
\end{align}

\subsection{Optimization algorithm}

In the following, we report a  numerical optimization algorithm in order to minimize the cost function $J$ in \eqref{eq:optimization_problem}  w.r.t. the parameters $\{\hatx, \swseq, \NNparx, \NNpary \}$. To this end, we employ the \emph{coordinate-descent} approach as described in  Algorithm~\ref{algo:coordinate_descent}. With a slight  abuse of notation, the optimization variable  $\hatx$ in Algorithm \ref{algo:coordinate_descent} denotes the finite-dimensional representation of the state signal $\hatx$, i.e.,  $\hatx=\{\hatx(t_0), \ldots, \hatx(t_{\nsamp-1})\}$.

\begin{algorithm}[!tb]
	\caption{Block coordinate descent for the estimation of states $\hatx$, mode sequence $\swseq$ and model parameter matrices $\NNparx, \NNpary$.} \label{algo:coordinate_descent} \vspace*{.1cm}
	\textbf{Input}: Training dataset $\mathcal{D}= \{\umeas(t_k),\ymeas(t_k)\}_{k=0}^{N-1}$;  initial guess  $\hatx^{(0)}, \swseq^{(0)}$;  tuning parameter $\alpha, \beta$; tolerance $\epsilon$, maximum number of iterations $n_{\mathrm{max}}$.  
	\vspace*{.2cm}\hrule\vspace*{.1cm}
	\begin{enumerate}[label=\arabic*., ref=\theenumi{}]
		\item \textbf{Iterate for} $n=1,\ldots$
		\begin{enumerate}[label=\theenumi{}.\arabic*., ref=\theenumi{}.\theenumii{}] 
			\item $ \NNparx^{(n)}, \NNpary^{(n)} \leftarrow \mathrm{arg} \underset{\NNparx,\NNpary}{\text{min}} J(\hatx^{(n-1)}, \swseq^{(n-1)}, \NNparx,\NNpary)$
			\label{algo:s1}
			\item $ \swseq^{(n)} \leftarrow \mathrm{arg} \underset{\swseq}{\text{min}} \ J(\hatx^{(n-1)}, \swseq, \NNparx^{(n)}, \NNpary^{(n)})$
			\label{algo:s2}
			\item $ \hatx^{(n)} \leftarrow \mathrm{arg}  \underset{\hatx}{\text{min}} \ \ J(\hatx, \swseq^{(n)}, \NNparx^{(n)},\NNpary^{(n)})$
			\label{algo:s3}
		\end{enumerate} 
		
		\item \textbf{Until}  $ \| J^{(n)} - J^{(n-1)} \| \leq \epsilon$ or $n= n_\mathrm{max}$ \label{algo:s4}
	\end{enumerate}
	\vspace*{.1cm}\hrule\vspace*{.1cm}
	~~\textbf{Output}: Estimated  $\{ \hatx(t_k)\}_{k=0}^{N-1}$, $\swseq$ and $\NNparx, \NNpary$.%
\end{algorithm}

Given an initial guess  $\hatx^{(0)}$ of the state and $\swseq^{(0)}$ of the mode  sequence, at each iteration $n \geq 1$, Algorithm~\ref{algo:coordinate_descent} alternates between three steps: Step~ \ref{algo:s1}, Step~\ref{algo:s2} and Step~\ref{algo:s3}. In particular, at Step~\ref{algo:s1},  model parameters $\NNparx, \NNpary$ are computed by solving \eqref{eq:optimization_problem} for  a \emph{fixed} state  $\hatx^{(n-1)}$ and mode sequence $\swseq^{(n-1)}$ obtained at the  iteration $(n-1)$. At  Step~\ref{algo:s2}, the mode sequence is estimated   for \emph{fixed} states $\hatx^{(n-1)}$ and fixed model parameters $\NNparx^{(n)}$ and $\NNpary^{(n)}$ obtained from Step~\ref{algo:s1} at the $n$-th iteration. Subsequently, at Step~\ref{algo:s3}, the state sequence $\hatx^{(n)}$ is estimated by minimizing the cost~\eqref{eq:optimization_problem} for \emph{fixed} model parameters $\NNparx^{(n)}$ and $\NNpary^{(n)}$ obtained from Step~\ref{algo:s1} and mode sequence $\swseq^{(n)}$ computed from Step~\ref{algo:s2}.
The procedure continues until a maximum number of iterations is reached, or a certain convergence criterion is met (Step \ref{algo:s4}).

\begin{Remark}\label{rem:initialization}
	Since the underlying optimization problem is non-convex,  convergence of Algorithm~\ref{algo:coordinate_descent} to the global optimal is sensitive to the initial guesses for states $\hatx^{(0)}$ and mode sequence  $\swseq^{(0)}$. A possible choice to initialize the state sequence  is to first identify a continuous-time LTI state-space model and set $\hatx^{(0)}$ to the states of the LTI model with small additive perturbations, \emph{i.e.,} $\hatx^{(0)}= \hatx_{\mathrm{LTI}} + v_x$ where, $v_x \sim \mathcal{N}(0, \sigma^2_x)$ with variance $\sigma^2_x$ chosen by the user. We remark that,  in practice,  Algorithm~\ref{algo:coordinate_descent}  can be run multiple times with different initial conditions and then choosing the best model parameters according to a figure of merit.        
\end{Remark}

\begin{Remark}\label{rem:convergence}
	We stress that in Algorithm~\ref{algo:coordinate_descent}, Steps~\ref{algo:s1} and \ref{algo:s3} can be solved \emph{analytically} via ordinary least squares, while Step~\ref{algo:s2} is solved to global optimality via dynamic programming. Thus, each sub-problem to be optimized within an iteration of the block corordinte descent is solved exactly to its unique optimal solution, which can be utilized to prove the convergence guarantees, see~\citep{bcd1, bcd2}. 
\end{Remark}

In the following section, we detail each step of the coordinate descent algorithm. Without loss of generality, for brevity, we set $\hC_{i} =  \hC$, $\hD_i = 0$ for all $i=1,\ldots,K$.  

\subsubsection{Step~\ref{algo:s1}: Optimization over model parameters $\NNparx, \NNpary$}

For a fixed mode sequence $\swseq$ and a fixed state sequence $\hatx$, the cost  function $J(\hatx^{(n-1)}, \swseq^{(n-1)}, \NNparx,\NNpary)$ in \eqref{eqn:J} can be optimized over the unknown model parameters $\NNparx,\NNpary$. This leads to a \emph{least-squares} problem described as follows.

Let $\theta_i = \left[\mathrm{vec}(\hA_i)^{\top} \ \mathrm{vec}(
\hB_i)^{\top} \right]^{\top} \in \mathbb{R}^{n_{\theta}}$, with $n_{\theta} = \nx(\nx+\nin)$ and let us define the  matrix $\Phi(t_j) \in \mathbb{R}^{n_{\theta} \times \nx}$ as follows
\begin{equation}
	\Phi(t_j) = \Delta t\smallmat{ \hatx(t_j) \\  \umeas(t_j)}\otimes I_{n_x}
\end{equation}
with $\otimes$ denoting the Kronecker  product.

The approximated state evolution eq.~\eqref{eqn:state-approx-2} can be written as,
\begin{align} \label{eqn:state-approx-4}
	\hatxI(t_k) &\approx \! \hat{\bf{x}}(0) \!+\! \sum_{j=0}^{k-1}   \Phi^{\top}(t_j) \theta_{\swest(t_j)}  
\end{align} 
Let us define the matrix $P \in \mathbb{R}^{(N-1) \times K}$  such that, for each of its row $i=1,\ldots, N-1$, the $j$-th column is set to $1$ if the active mode at time $t_i$ is $\swest(t_i) = j$, \emph{i.e.}, the $(i,j)$-th entry  $P_{i,j}$ is defined as
\begin{align*}
	P_{i,j} &= 1  \ \ \mathrm{if} \ \swest(t_i) = j \\
	& = 0 \ \ \mathrm{otherwise}
\end{align*}
and let us define $\bar{P} = P \otimes I_{n_{\theta}} \in \mathbb{R}^{(N-1)n_{\theta}\times Kn_{\theta}}$.

With the matrices  defined above, the relation~\eqref{eqn:state-approx-4} can be written in the matrix form as follows:

\begin{align}
	&\underset{ \Delta \hatx}{ \underbrace{	\smallmat{
				\hatxI(t_1) - \hatx(0)\\
				\hatxI(t_2)-\hatx(0)\\
				\vdots \\
				\hatxI(t_{N-1})-\hatx(0)
	}}} = \\ &  \underset{ \Psi} {\underbrace{ \smallmat{
				\Phi^{\top}(t_0) & 0 & 0 & \cdots  &0 \\
				\Phi^{\top}(t_0)& 	 \Phi^{\top}(t_1) & 0 & \cdots  & 0 \\
				\vdots & & & &  \\
				\Phi^{\top}(t_0)&  \Phi^{\top}(t_1)&  \Phi^{\top}(t_2)  & \cdots  & \Phi^{\top}(t_{N-2})
	}}} \bar{P} \smallmat{
		\theta_{1} \\
		\theta_{2}\\
		\vdots \\
		\theta_{K} 
	} 
\end{align}

Based on the above definitions, the cost function \eqref{eqn:J} can be  re-written as
\begin{align} \label{eqn:Jnew_1}
	J= \underset{J_{\bf y}}{\underbrace{  \left\|   \tilde{C} \hatx - \ymeas \right\|^2} } 
	+
	\alpha \underset{J_{\bf x}}{\underbrace{ \left\|  \Delta \hatx - \Psi\bar{P} \Theta \right\|^2 \; \Delta t}}
\end{align}
where $\tilde C = blk(\hC)$ is a block-diagonal matrix, $\hatx $ and $\ymeas$ are the sequences of estimated states and measured outputs respectively. 
Note that, for a given state estimates $\hatx$ and a given mode sequence $\swseq$, the matrices $\Psi$ and $\bar{P}$ can be pre-computed and  
thus, \eqref{eqn:Jnew_1} is a \emph{least-squares} problem in the unknown model parameters $\{ \theta_i\}_{i=1}^{K}, \tilde{C}$ (\emph{i.e.}, $\{\hA_i,\hB_i\}_{i=1}^{K}, \hC$), which can be solved analytically. 

\subsubsection{Step~\ref{algo:s2}: Optimization over mode sequence $\swseq$}

Given the estimates of the model parameters $\{\hA_i,\hB_i\}_{i=1}^{K}, \hC$ computed at Step~\ref{algo:s1} and given a fixed state sequence $\hatx$, the cost  function $J(\hatx^{(n-1)}, \swseq, \NNparx^{(n)},\NNpary^{(n)})$ in \eqref{eqn:J} can be optimized over the unknown mode sequence $\swseq$ via discrete \emph{Dynamic Programming} (DP) algorithm \citep{BPBB2018}. 

The DP algorithm to estimate the mode sequence $\swseq$ is summarized as follows.  Let $\ell(\hatx(t_k),\umeas(t_k),  \theta_{\swest(t_k)})$ be the transition cost defined as 
\begin{align*}
	&\ell( \hatx(t_k), \umeas(t_k), \theta_{\swest(t_k)})\\  &=  \left\|  \hatx(t_{k+1}) - \hatx(t_k)- \Delta t \hA_{\swest(t_k)}\hatx(t_k)- \Delta t\hB_{\swest(t_k)}\umeas(t_k) \right\|^2
\end{align*}

We compute a  matrix of cost $V \in \mathbb{R}^{K \times (N+1)}$ and a matrix of indices $U \in \mathbb{R}^{K \times N}$ as follows:

First, the terminal cost $V_{i, N} \in \mathbb{R}$ is computed for all modes  
\begin{align*}
	V_{i, N-1} =  \LcostMode(i) + 		\ell( \hatx(t_{N-1}), \umeas(t_{N-1}), \theta_{i}) \ \ i=1,\ldots,\nmodes.
\end{align*}

Next, the cost $V_{i,k}$ and indexes $U_{i,k}$ at time $t_k$ are computed with following dynamic programming recursions, backwards in time for $k=  N-2, \ldots ,1, 0$
\begin{align*}
	U(i,k)  & = \underset{j=1,\ldots,\nmodes}{\mathrm{arg} \  \mathrm{min}}  \ \ \{ V_{j, k+1} + \LcostTrans(j, i) \},  \quad i=1,\ldots, K \\
	V_{i, k} &= \LcostMode(i) + \ell( \hatx(t_{k}), \umeas(t_{k}), \theta_{i}) +   V_{ U(i,k), k+1} \\&+ \LcostTrans(U(i,k), i), \\
	V_{i, 0} &= \LcostInit(i) + \underset{j=1,\ldots,\nmodes}{ \mathrm{min}} \ \ \{V_{j, 1} + \LcostTrans(j,i) \} 
\end{align*}

The minimum cost mode sequence $\swseq$ is retrieved forward in time by setting
\begin{align*}
	\swest_{t_{0}} &=   \underset{j=1,\ldots,\nmodes}{\mathrm{arg} \  \mathrm{min}} \ \ V_{j, 0} \\ 
	\swest_{t_{k}} &=  U(\swest_{t_{k-1}} ,k)  \ \ \ k= 1,\ldots, N-1
\end{align*}

\subsubsection{Step~\ref{algo:s3}: Optimization over states $\hatx$ }

By assuming that the initial conditions $\hatxI(t_0)$ and $\hatx(t_0)$ are equal, i.e.,
\begin{align}
	\hatxI(t_0) = \hatx(t_0),
\end{align}
the relation~\eqref{eqn:state-approx-3} can be written in the matrix form:
\begin{align}
	&	\smallmat{\hatxI(t_0) \\
		\hatxI(t_1)\\
		\hatxI(t_2)\\
		\vdots \\
		\hatxI(t_{N-1})} =  \nonumber \\
	& \underset{ \tilde A} {\underbrace{\smallmat{	I & 0 & 0 & \cdots & 0 &0 \\
				I + \Delta t \hA_{\swest(t_0)}& 0 & 0 & \cdots & 0 & 0 \\
				I + \Delta t \hA_{\swest(t_0)}& \Delta t \hA_{\swest(t_1)} & 0 & \cdots & 0 & 0 \\
				\vdots & & & & & \\
				I + \Delta t \hA_{\swest(t_0)}& \Delta t \hA_{\swest(t_1)} & \Delta t \hA_{\swest(t_2)}  & \cdots & \Delta t \hA_{\swest(t_{N-2})} & 0} }} \smallmat{\hatx(t_0) \\
		\hatx(t_1)\\
		\hatx(t_2)\\
		\vdots \\
		\hatx(t_{N-1})} \nonumber \\
	+ & \underset{ \tilde B} {\underbrace{\smallmat{
				0  \\
				\Delta t \hB_{\swest(t_0)}\umeas(t_0)\\
				\Delta t \hB_{\swest(t_0)}\umeas(t_0) + \Delta t \hB_{\swest(t_1)}\umeas(t_1)\\
				\vdots \\
				\Delta t \hB_{\swest(t_0)}\umeas(t_0) + \Delta t \hB_{\swest(t_1)}\umeas(t_1) + \ldots + \Delta t \hB_{\swest(t_{N-2})}\umeas(t_{N-2})
	}}}
\end{align}

Thus, based on the above approximation, the cost function \eqref{eqn:J} can be also re-written as
\begin{align} \label{eqn:Jnew}
	J= \underset{J_{\bf y}}{\underbrace{  \left\|   \tilde{C}\hatx- \ymeas \right\|^2} } 
	+
	\alpha \underset{J_{\bf x}}{\underbrace{ \left\|  (\tilde A - I) \hatx + \tilde B \right\|^2 \; \Delta t}}
	+ \Lcost(\swseq),
\end{align}
where $\tilde C = blk(C)$, which can be solved for $\hatx$ via ordinary least-squares.

\section{SIMULATION EXAMPLE}\label{sec:CaseStudy}

The performance of the proposed algorithm is assessed via a simulation case study. All computations are carried out on an i7 1.9-GHz Intel core processor with 32
GB of RAM running MATLAB R2019a.

We consider a continuous-time linear switched system governed by \eqref{eq:init} having $\nmodes= 2$ modes with subsystem matrices given as follows~\citep{goudjil2020ecc}:
\begin{align*}\label{eqn:sys_eg1}
	&\left[
	\begin{array}{c|c}
		A_1  &  B_1   \\
		\hline
	C_1 & D_{1}  \\ 
	\end{array}
	\right]  =  \left[
	\begin{array}{c c | c }
	0  &-120  &120    \\
		1 & -4 & -12   \\
		\hline
		0   &1    &     0 
	\end{array}
	\right],\\
	&\left[
	\begin{array}{c|c}
		A_2  &  B_2   \\
		\hline
		C_2 &  D_{2}  \\ 
	\end{array}
	\right]  =   \left[
	\begin{array}{c c | c }
		0  &-50  &53    \\
		1 & -1.8 & 25  \\
		\hline
		0   &1    &     0 
	\end{array}
	\right]
\end{align*} 
The system is excited with a zero-mean  Gaussian input signal having unit variance, $\umeas(t) \sim \mathcal{N}(0,1)$. The dynamics switches between the two subsystems with a Markov switching signal, such that the true mode $\sw_{t_{k}}$ has $\pi = 10 \%$ probability of being different from  $\sw_{t_{k-1}}$, starting from $\sw_{t_{0}} = 1$. The system  belongs to a class of switched models termed as continuous-time  jump Markov linear systems~\citep{costa20136mjls}. 
Training dataset of $N = 400$ samples is gathered, sampling the output and input trajectories with a sampling time of $\Delta t = 0.01$ s. The output is corrupted by an additive white Gaussian noise $\ymeas(t_k) = \yo(t_k) + \eta(t_k)$ where $ \eta(t_k) \sim \mathcal{N}(0,\sigma_{\eta}^2)$ with $\sigma_{\eta} = 0.025$, which corresponds to signal-to-noise ratio of $30$ dB.

For identification, we consider an LSS model structure \eqref{eq:final} with state dimension set to the true system dimension $\nx =2$ and number of modes set to $\nmodes=2$. The model matrices $\NNparx$, $\NNpary$, the mode sequence $\swseq$ and the state sequence $\hatx$ are estimated by running the coordinate descent Algorithm~\ref{algo:coordinate_descent} for $n_{\max}= 1000$ iterations. The average computational  time for each iteration of the algorithm is $0.1$ s, which includes the time to compute the model matrices and states via ordinary least-squares and estimation of mode sequence via dynamic programming recursions. In total, the entire identification problem is completed in about 500~s, with 5 different initial guesses.  

\begin{figure}[!t]
	\centering
	\includegraphics[scale=0.7]{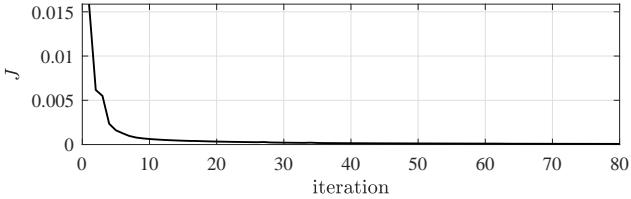}
	\caption{Training: Cost function \emph{vs} number of iterations.}
	\label{fig:cost}
\end{figure}

To asses the convergence properties,  cost $J$  is plotted in Fig.~\ref{fig:cost} against the iterations of the coordinate descent algorithm.  

As mentioned in Remark~\ref{rem:initialization},  the initial guess $\hatx^{(0)}$ for the state sequence is set to the states of an identified CT LTI state-space model\footnote{The CT LTI model is identified using MATLAB's system identification toolbox with command $\mathrm{ssest}$ which employs $\mathrm{N4SID}$ subspace algorithm.} with small additive perturbations, \emph{i.e.,} $\hatx^{(0)}= \hatx_{\mathrm{LTI}} + v_x$ where, $v_x \sim \mathcal{N}(0, \sigma^2_x I)$ with  $\sigma_x = 0.01$. 
The initial guess for the mode sequence $\swseq^{(0)}$ is chosen randomly.

For the mode loss  $\Lcost(\swseq)$ (see eq.~\eqref{eq:mode_loss}), we set initial mode cost $\LcostInit(\sw_{t_{0}}) = 0$, mode cost $\LcostMode(\sw_{t_{k}}) = 0$. The transition mode cost $\LcostTrans$ is chosen as follows:
\begin{equation}\label{eq:transition loss}
	\LcostTrans(\sw_{t_{k}}, \sw_{t_{k-1}}) = \left\{\begin{matrix}
		- \tau \log(1-\pi) \quad \mathrm{if} \  \sw_{t_{k}} = \sw_{t_{k-1}} \\ 
		\ \tau \log(\pi)   \quad \quad \quad \ \mathrm{if} \  \sw_{t_{k}} \neq \sw_{t_{k-1}}
	\end{matrix}\right.	
\end{equation} 
with $\pi = 0.1$ and $\tau = 10^{-6}$. The regularization hyper-parameter $\alpha$ is set to $0.01$. The hyper-parameters $\tau$ and $\alpha$ are chosen via a grid search.

\begin{table}[tb!]
	\caption{True \emph{vs} estimated transfer functions for mode $1$ and mode $2$.}
	\label{Tab:tf}
	\vspace{-0.2cm}
	\begin{center}
		\begin{tabular}{|c|c|c|}
			\hline
			Mode	&True & Estimated  \\
			
			\hline
			& &  \\
			$G_{1}(s)$ 	& $\frac{-12s+120}{s^2+4s+120}$ & $\frac{-12.01s+122.3}{s^2+3.98s+119}$\\
			& &  \\
			\hline
			& &  \\
			$G_{2}(s)$	 & $\frac{25s+53}{s^2+1.8s+50}$&	$\frac{24.97s+52.9}{s^2+1.89s+49.3}$ \\ 
			& &  \\
			\hline
		\end{tabular}
	\end{center}
\end{table} 

\begin{figure}[!t]
	\centering
	\includegraphics[scale=0.8]{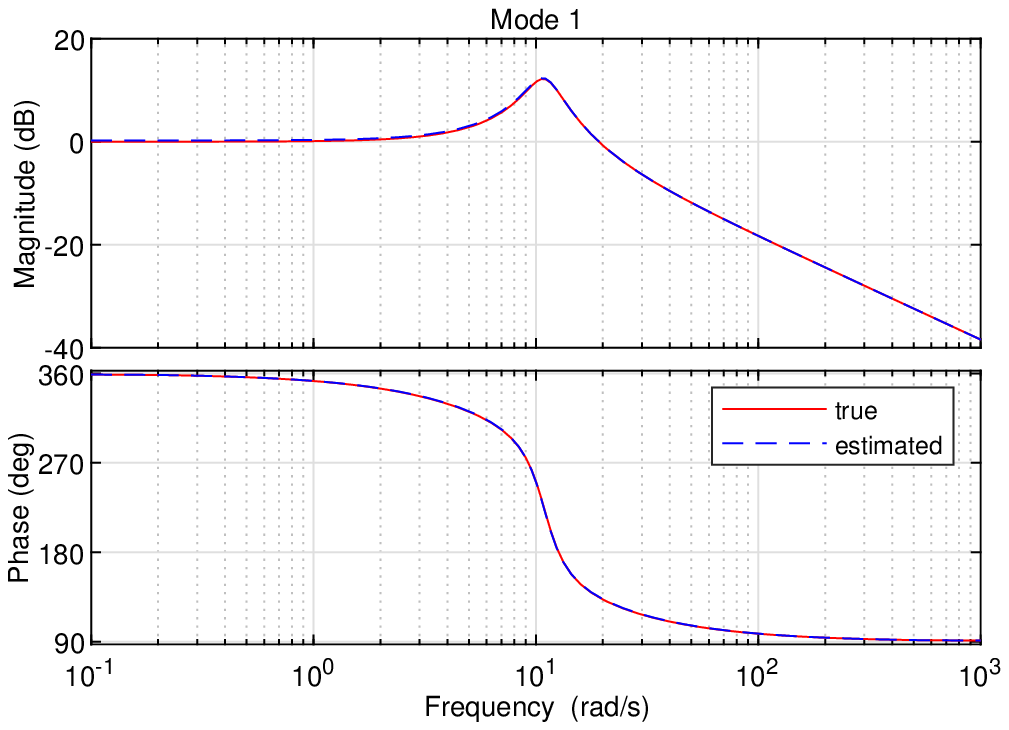}
	\ 
	\
	\includegraphics[scale=0.8]{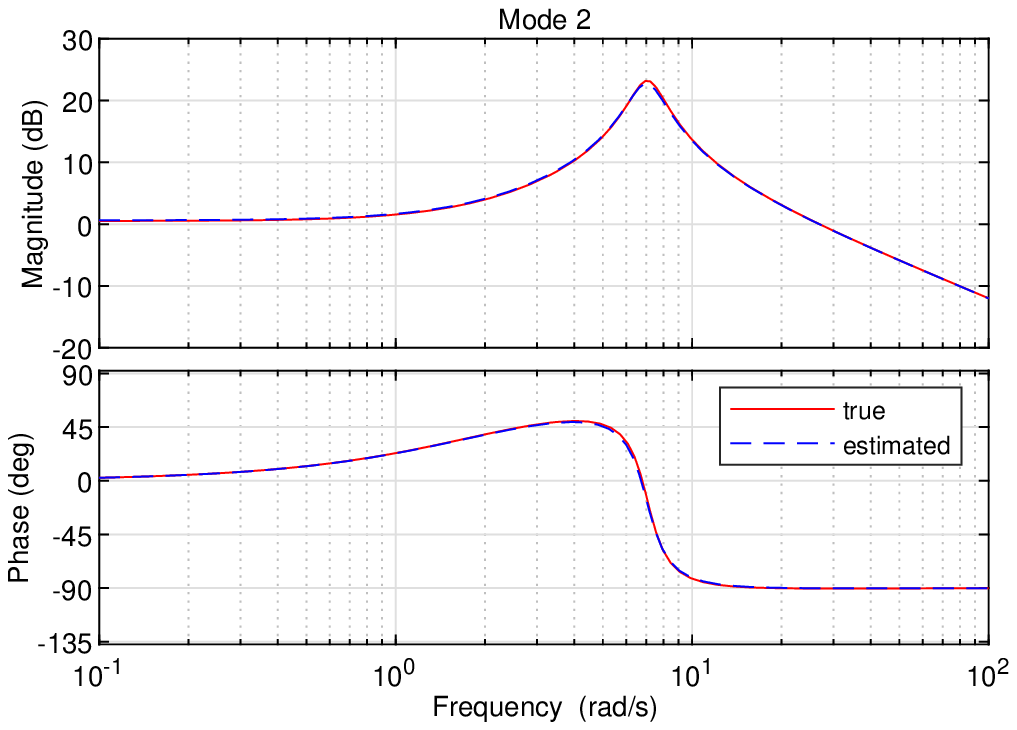}
	
	\caption{Bode plot: true (red) \emph{vs} estimated (dashed blue) model for mode $1$ (top panel) and mode $2$ (bottom panel).}
	\label{fig:bode}
\end{figure}

The true and estimated transfer functions $G(s)$ of the two linear subsystems are reported in Table~\ref{Tab:tf}. The corresponding Bode plots of the  subsystems are depicted in Fig.~\ref{fig:bode}. The obtained results show that the model parameters of each subsystem have been identified with high accuracy and input-output behavior of the estimated  linear subsystems matches closely to that of the true subsystems.

\begin{figure}[!t]
	\centering
	\includegraphics[scale=0.65]{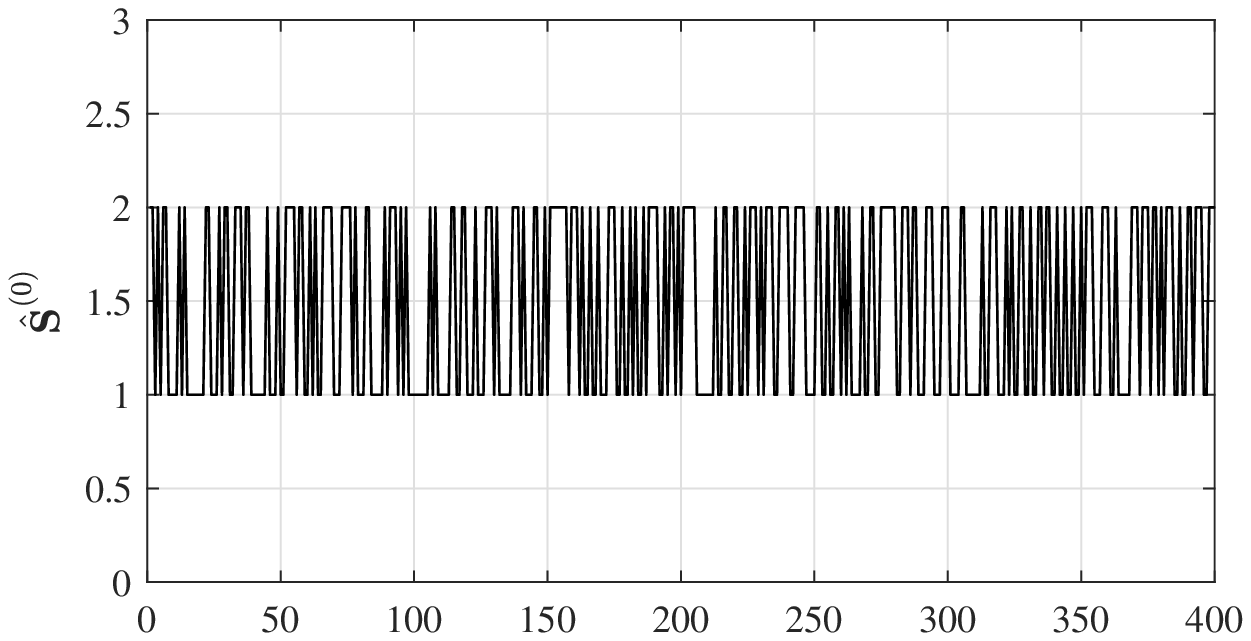}
	\includegraphics[scale=0.65]{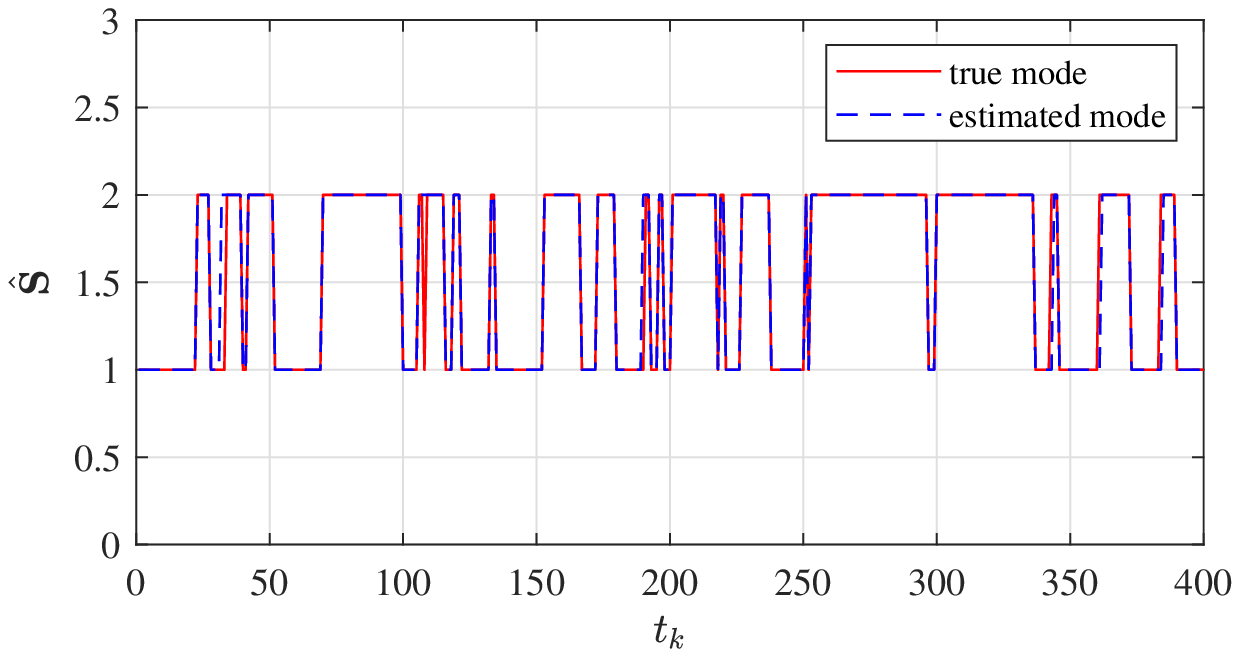}
	\caption{Estimation of mode sequence: Top panel: Random intial guess $\swseq^{(0)}$, Bottom panel: True (red) \emph{vs} estimated (blue) mode sequence with mode fit percent = $98.25 \%$.}
	\label{fig:mode}
\end{figure}

The performance of the proposed identification algorithm is further assessed  in terms of mode sequence estimation,
quantified via a \emph{mode fit} (MF) index $\frac{100}{N}\sum_{k=0}^{N-1} \delta(\swest_{t_k}, \sw_{t_{k}})$ where $\delta(i,j)$ is the Kronecker delta function.  Fig.~\ref{fig:mode} shows the true \emph{vs} the estimated mode sequence.  Only $7$ out of $400$ modes have been incorrectly classified which leads to a mode fit of $98.25 \%$. It is clear from Fig.~\ref{fig:mode} that, starting from a random initial mode sequence, the proposed algorithm is able to reconstruct the true mode sequence accurately.

\begin{figure}[!t]
	\centering
	\includegraphics[scale=0.6]{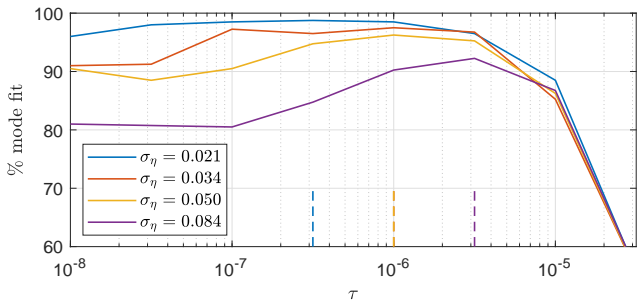}
	\caption{Mode fit percent \emph{vs} $\tau$ for different noise levels. Optimal value of $\tau$ shown with dashed lines.}
	\label{fig:tau}
\end{figure}

Furthermore, we asses the effect of varying noise levels $\sigma_{\eta}$ as well as regularization hyper-parameters $\tau$ on the performance of the proposed algorithm quantified via mode-fit index. The results are summarized in Fig.~\ref{fig:tau}, which shows the percentage of correctly classified modes against the values of $\tau$ for different noise standard deviations $\sigma_{\eta} = \{0.02,0.03,0.05,0.08\}$ corresponding to signal-to-noise ratios $\{32, 28,24,20 \}$ dB, respectively. From the choice of the mode transition loss $\LcostTrans$ in \eqref{eq:transition loss}, higher values of $\tau$ implies more penalty on the change of mode. In other words, for large values of $\tau$, mode change is discouraged and only single mode is recognized (typically, the value of the initial mode $\swest_{t_{0}}$ is retained), leading to a lower mode fit percent as seen in Fig.~\ref{fig:tau}. The hyperparameter $\tau$ and the choice of  $\LcostTrans$ thus act as a tuning knob, which can be chosen via cross-validation,  depending upon either fast or slow switching dynamics.

Finally, in order to analyze the statistical properties and robustness of the proposed algorithm, we perform a Monte-Carlo (MC) analyses with $50$ MC runs. At each MC run, data is gathered by exciting the system  with a new realization of the input, switching signal and noise. The variance of the noise distribution is set such that the average SNR for each run is $30$ dB. 
Algorithm~\ref{algo:coordinate_descent} is run for $5$ different initial guesses for the states  $\hatx^{(0)}$  and mode sequence $ \swseq^{(0)}$, with $n_{\mathrm{max}}=1000$ iterations setting $(\alpha, \tau)= (0.01,3\cdot10^{-7})$. Among  $5$  different initializations,  model parameters obtained from the run having maximum \emph{best fit rate}:  $\mathrm{BFR} = 100\left( 1-\sqrt{\frac{\sum_{k=0}^{N-1}(\haty(t_k) -\ymeas(t_k))^2}{\sum_{k=0}^{N-1}(\ymeas(t_k)-\bar{\ymeas})^2} } \right)  \%$ are chosen.

\begin{figure}[!t]
	\centering
\includegraphics[scale=0.6]{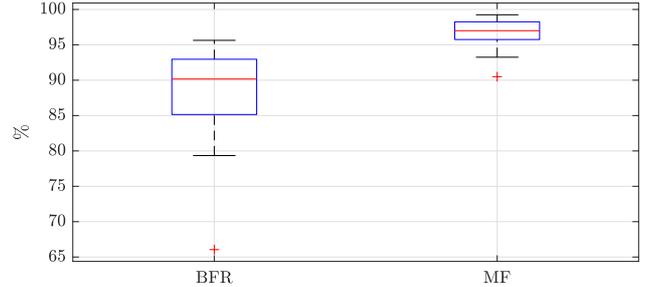}
	\caption{Monte-Carlo analysis: Best fit rate and mode fit percentage over $50$ MC runs.}
	\label{fig:BFR_MF}
\end{figure} 

The box-plots of the mode fit (MF) and BFR indexes over $50$ Monte-Carlo runs are shown in Fig.~\ref{fig:BFR_MF}.  
We  observe that satisfactory performance is obtained in terms of reconstruction of the output as well as the mode sequence.
We remark that, although  convergence is not guaranteed for every run of the algorithm, in practice,  running the algorithm with $5$ different initializations was sufficient to achieve  accurate model parameter estimates.

\section{CONCLUSIONS}
In this work, we have presented an integral architecture for  continuous-time identification of  switched state-space models.  The proposed approach can be seen as the first step towards developing a generic framework for direct  identification of continuous-time state-space hybrid dynamical systems. The  presented analysis has shown that satisfactory results have been achieved for identifying a  Markov jump linear system, in terms of reconstruction of the mode sequence as well as  capturing the input-output behaviours of the linear submodels. 
Future works will focus on developing refinement strategies in order to  improve the estimation of mode sequence and robustness w.r.t. to initial conditions.


\bibliography{references}
                                                   







\end{document}